\documentclass[fleqn, 9pt, twocolumn]{wlscirep}
\usepackage[utf8]{inputenc}
\usepackage[T1]{fontenc}
\usepackage{mathtools}
\usepackage{amsmath,amssymb, bm}
\usepackage[font=small,labelfont=bf,
   justification=justified,
   format=plain, figurename=Figure]{caption}
\providecommand\JournalTitle[1]{#1}
\allowdisplaybreaks
\usepackage{color}
\providecommand{\diff}[1]{{#1}} 

\title{Emergence of Kinship Structures and Descent Systems: Multi-level Evolutionary Simulation and Empirical Data Analysis}
\author[1]{Kenji Itao}
\author[1, 2, *]{Kunihiko Kaneko} 

\affil[1]{Department of Basic Science, Graduate School of Arts and Sciences, University of Tokyo, Komaba 3-8-1, Meguro-ku, Tokyo 153-8902, Japan.}
\affil[2]{Research Center for Complex Systems Biology, University of Tokyo, Komaba 3-8-1, Meguro-ku, Tokyo 153-8902, Japan.}

\affil[*]{kaneko@complex.c.u-tokyo.ac.jp}


\begin{abstract}
In many indigenous societies, people are categorised into several cultural groups, or clans, within which they believe to share ancestors. Clan attributions provide certain rules for marriage and descent. Such rules between clans constitute kinship structures. Anthropologists have revealed several kinship structures. Here, we propose an agent-based model of indigenous societies to reveal the evolution of kinship structures. In the model, several societies compete. Societies themselves comprise multiple families with parameters for cultural traits and mate preferences. These values determine with whom each family cooperates and competes and are transmitted to a new generation with mutation. The growth rate of each family is determined by the number of cooperators and competitors. Through this multi-level evolution, family traits and preferences diverge to form clusters that can be regarded as clans. Subsequently, kinship structures emerge, including dual organisation and generalised or restricted exchange, as well as patrilineal, matrilineal, and double descent systems. These structures emerge depending on the necessity of cooperation and the strength of mating competition. Their dependence is also estimated analytically. Finally, statistical analysis using the Standard Cross-Cultural Sample, a global ethnographic database, empirically verified theoretical results. Such collaboration between theoretical and empirical approaches will unveil universal features in anthropology.
\end{abstract}

\begin{document}

\flushbottom
\maketitle

\thispagestyle{empty}

\section*{Introduction}
\diff{Marriage and descent form the basic units of society; that is, families. Kinship relationship stipulates the alliance of families and organises social structures.
It is considered as one of the oldest, most frequent human social organisations \cite{levi1969elementary, service1962primitive}.}
In various indigenous societies, people constitute a cultural association, or clan, in which they are culturally (but not necessarily biologically) related \cite{fox1983kinship, levi1969elementary, maddock1969alliance}. Marriage and descent relationships are, thus, determined by clan attributions \diff{(i.e. the clan to which individuals belong)}. Specifically, marriage within a clan is often prohibited by the symbolic incest taboo \cite{levi1969elementary, leach1954political, malinowski1963sex, murdock1949social, hopkins1980brother, Hill2011}. The rule can further specify the clan from which one must select a mate from and that to which children must belong \cite{levi1969elementary}. 
\diff{Kinship relationships also regulate social relationships, such as cooperation or rivalry \cite{fox1983kinship}.
The elucidation of kinship systems has been a core theme in cultural and evolutionary anthropology \cite{fox1983kinship, shenk2011rebirth}. Anthropologists have characterised kinship systems by focusing on the affinal network of clans, namely, kinship structure \cite{levi1969elementary}; or by
the categorisation of relatives by ego, namely, kinship terminology \cite{murdock1949social, passmore2021kin}. Here, we consider kinship structures.}

\diff{Kinship structures are diverse yet patterned. They can be classified into several types, according to the length of cycles composed by the marriage and descent relationships of clans \cite{levi1969elementary, white1963anatomy}. 
For example, if a rule exists for women in clan X to marry men in clan Y, the marriage relationship is represented by $X \Rightarrow Y$. 
If everyone can potentially have mates, the relationships of clans should be $X \Rightarrow Y \Rightarrow \cdots \Rightarrow X$. Here, marriage relationships of clans constitute a cycle, the length of which is termed marriage cycle $C_m$ (e.g., $C_m = 3$ if $X \Rightarrow Y \Rightarrow Z \Rightarrow X$). Similarly, if the children belong to clan B, and their father to clan A, it represents the descent relationship $A \rightarrow B$. This relationship also constitutes the cycle, and its length is termed the descent cycle $C_d$.
Notably, a clan is not always a residence group. Family members of different generations can have different clan attributions \cite{romney1958simplified}.
For example, when children inherit their father's surname but live in their mother's location, the children's attribution, determined by both surname and location, differs from those of both their father and mother. 
(This can also be regarded as children belonging to several associations following each parent simultaneously \cite{service1962primitive}.)}

Kinship structures are characterised by marriage cycle $C_m$ and descent cycle $C_d$.
The classes include the incest structure -- conducting endogamy without the symbolic incest taboo ($C_m = C_d = 1$, i.e. without division of clans); dual organisation -- a direct exchange of brides between two clans ($C_m = 2, C_d = 1$); generalised exchange -- an indirect exchange of brides among more than two clans ($C_m \ge 3, C_d = 1$); and restricted exchange -- a direct exchange of brides with the flow of children to different clans ($C_m = C_d = 2$). Structures with $C_m \ge 3$ and $C_d \ge 2$ are rarely observed.

\diff{In this paper, we discuss the evolution of three types of descent systems.
When children belong to the same clan as either their father (or mother), the descent system is classified as patrilineal (or matrilineal) descent, respectively. In these cases, $C_d = 1$, and paternally (or maternally) inherited trait is significant for characterising clans.
Conversely, when $C_d > 1$, children inherit cultural traits from both parents independently and have clan attributions different from either parent. When both paternally and maternally inherited traits are significant for characterising clans, the system is termed double descent. In the above cases, traits are assumed to be independently inherited through paternal and maternal lines. (In some societies, however, people can choose either their father's or mother's traits to inherit in each generation (ambilineal descent) or they concern genealogical distance only (bilateral descent) which exceed the scope of our model \cite{levi1965future, service1962primitive}.)}

Ethnographic reports provide examples of various descent systems \cite{levi1969elementary, double_descent, Goody1961}. Global data indicate that patrilineal descent is dominant over matrilineal or double descent \cite{murdock1969standard}.
Evolutionary anthropologists attribute this imbalance to the higher investment efficiency of reproductive resources for sons than for daughters \cite{hartung1981paternity, holden2003matriliny, shenk2011rebirth}. 
However, this perspective ignores the distinction of cultural associations within societies regarding symbolic traits. Indeed, the identity of the categorical descent group is more emphasised than genetic relatedness in some cooperative actions \cite{alvard2003kinship, alvard2011genetic}.

\diff{Moreover, cultural traits of families and kinship are slow to change, because they are inherited in families and regulated by social norms \cite{cavalli1981cultural}. 
Empirical studies confirm such slow changes \cite{guglielmino1995cultural, mulder2001study, minocher2019explaining}. 
To consider the inheritance of family traits from parents or their relatives with slight changes, it is appropriate to model their long-term evolution through the accumulation of small variations, as represented by mutations.
Notably, families constitute society, whereas society provides the environment for families.
Consequently, we adopted a framework involving the multi-level evolution of families and societies.
Multi-level evolution is a framework generally applied for discussing the evolution of group-level structures in hierarchical systems \cite{traulsen2006evolution, takeuchi2017origin, spencer2001multilevel, turchin2009evolution}.
In this study, we aimed to reveal the emergence of various kinship structures and descent systems from family interactions depending on environmental conditions.}

\diff{We thus modelled the family behaviour in indigenous societies. In the model, evolution is considered at two levels: that of the family, which is an individual agent of the model; and that of society, which is a group of families.
We assigned each family a trait $t$ and a mate preference $p$. 
Social relationships of families -- including cooperation, competition, and marriage -- are determined by their traits and preferences.
Families grow through interactions with other families, which subsequently leads to the growth of societies.
As a result of this multi-level evolutionary simulation, $t, p$ values of families diverge and form clusters within each society. These clusters are exogamous groups of families, which can be regarded as clans. By tracing the marriage and descent relationships of the emergent clans, we demonstrate the evolution of kinship structures and descent systems.
Previously, we have constructed an intricate model to illustrate the evolution of kinship structures \cite{itao2020evolution}.
Here, we introduce a simplified model suitable for studying the evolution of both kinship structures and descent systems, together with analytical estimates and empirical tests on a cross-cultural database.}

For data analysis, we used the global ethnographic database of premodern societies, the Standard Cross-Cultural Sample (SCCS) \cite{murdock1969standard, kirby2016d}. The SCCS contains 186 societies, considered culturally and linguistically independent of each other, \diff{(even if some correlation exists due to the shared ancestry in the strict sense \cite{minocher2019explaining}).} The data allowed us to quantitatively analyse cultural adaptations to environments \cite{marsh1967comparative, bernard2017research}.
Previous studies have investigated conditions that generally favour cousin marriages \cite{hoben2016factors} and polygamy \cite{white1988causes}. However, it is difficult to further explain the diversity in cousin marriage and kinship structures solely from correlation analyses \cite{racz2020social}. 
Thus, we demonstrate that the collaboration between theoretical simulation and statistical analysis can enable us to unveil the origins of, and conditions for, each kinship structure.

The remainder of this paper is organised as follows. In the next section, we introduce a simplified model. Then, using evolutionary simulations, we demonstrate the emergence of kinship structures and descent systems, and uncover the conditions for their emergence. We also estimate these conditions analytically. Next, by analysing the SCCS, the theoretical results are verified. Finally, we discuss how the present method, which combines theoretical models and empirical data analysis, is relevant to exploring anthropological phenomena.

\section*{Model}
\begin{figure}[tb]
\centering
    \includegraphics[width= 1.0\linewidth]{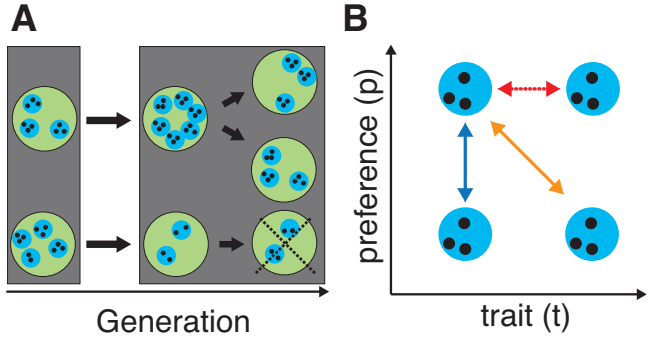}
  \caption{Schematic of the model.
(A) Life cycle of the model. Societies (green) consist of families (blue), whose population (black) grows. The grey frame represents a single generation. Families grow through interactions with other families in the same society.
When the population of a family (society) exceeds a given threshold, the family (society) splits. Subsequently, another society is removed from the system at random to keep the number of societies fixed. 
(B) Families cooperate with kin and mates (blue and orange solid lines), and conflict with rivals (red dashed line) depending on their traits $\bm{t}$ and mate preferences $\bm{p}$.
Families $i$ and $j$ are kin (blue) when $|\bm{t}^i - \bm{t}^j| / \tau$ is sufficiently small, mates (orange) when $|\bm{t}^i - \bm{p}^j| / \tau$ or $|\bm{p}^i - \bm{t}^j| / \tau$ is small, and rivals (red) when $|\bm{p}^i - \bm{p}^j| / \tau$ is small. Only the relationships with the upper-left family are shown. In the figure, we plotted the relationships in a two-dimensional space for simplicity. In the model, however, $\bm{t}$ and $\bm{p}$ are both two-dimensional. Thus, the relationships are considered in four-dimensional space.
}
\label{fig:scheme}
\end{figure}

The model is described below in general terms (see the Methods section for further details). Fig. \ref{fig:scheme} shows a schematic of our model. Families grow by interacting with other families in the same society (Fig. \ref{fig:scheme}(A)). 
\diff{Here, we ignore the explicit interaction between societies including migration, for simplicity. However, the following results are robust against slight migrations, as shown in Fig. S1.}
At the time of marriage, family members independently build new families of their own. The society splits in half when the number of families therein doubles its initial value $N_f$. At this time, another society is removed at random; thus, the number of societies in the entire system remains fixed at $N_s$. However, the number of families fluctuates between $0$ and $2 N_f$. This process introduces society-level selection, such that societies that grow at a faster rate replace others. This can be interpreted as invasion, imitation, or the coarse-grained description of a growing system. This framework, known as the multi-level selection, has been widely adopted in biological and social evolution studies to explain group-level structures \cite{itao2020evolution, itao2021evolution, Traulsen2006, spencer2001multilevel, takeuchi2017origin, Wilson2003, turchin2009evolution}.
\diff{Previously, we considered a model with three layers, including the intermediate layer of ``lineages'' between families and societies. Here, we simplify the model by eliminating it, to explore the generality of the results and to be suitable for analytical calculations \cite{itao2020evolution}.}

\diff{Moreover, each family has a pair of cultural traits and mate preferences that are culturally transmitted to the next generation. The traits can represent any social features by which people can measure their cultural similarity, for instance, surnames, occupations, or totems \cite{levi1962pensee}.
In the following section, we demonstrate that initially uniform traits gradually diverge to be discrete for distinguishing family groups.}
Marriage occurs when men's traits are close to women's preferences. In our model, this point is the sole asymmetry between men and women, consistent with 
anthropological studies stating that in most societies, brides' families determine whether grooms are suitable for marriage \cite{levi1969elementary, levi1962pensee}.

There are two pathways for cultural transmission: paternal and maternal. Hence, we require the two-dimensional trait $\bm{t} = (t_1, t_2)$ and preference $\bm{p} = (p_1, p_2)$. Thus, when a man in family $i$ and a woman in family $j$ are married, their children will have the trait $\bm{t} = (t_1^i, t_2^j)$ and preference $ \bm{p} = (p_1^i, p_2^j)$. 
At the time of cultural transmission, we add noise $\bm{\eta} = (\eta_1, \eta_2)$ to $\bm{t}$ and $\bm{p}$, independently sampled from a normal distribution with mean $0$ and variance $\mu^2$. Similar to genetic mutations in evolutionary biology, cultural traits are slightly modified when they are transmitted \cite{cavalli1981cultural}. Such cultural traits are used to categorise social groups, even without genetic relatedness \cite{sperber2004cognitive}.
\diff{Previously, we assumed that $t_1$ and $p_1$ are inherited from the father, and $t_2$ and $p_2$ are inherited either from the father or the mother, depending on the families' strategies \cite{itao2020evolution}. However, this assumption limits the evolution of descent systems, as the matrilineal descent system is set to be harder to evolve.
Here, we revised this to enable discussion on the evolution of various descent systems. (Notably, paternal and maternal traits are still supposed to be inherited independently. Hence, those descent systems in which both parents' traits are multiply referred to, exceed the scope of our model.)}

First, we introduced cooperative relationships with cultural kin and mates (blue and orange solid lines in Fig. \ref{fig:scheme}(B)). Families cooperate with those who have traits similar to their own, and those who prefer (or are preferred by) them.
\diff{In the model, the degree of cooperation between family $i$ and $j$ is given by $\exp(-\min(|\bm{t}^i-\bm{t}^j|, |\bm{t}^i-\bm{p}^j|, |\bm{p}^i-\bm{t}^j|) ^2/\tau^2)$, where $|\bm{t}^i-\bm{t}^j| = \sqrt{(t_1^i - t_1^j)^2 + (t_2^i - t_2^j)^2}$ represents Euclidean distance and $\tau$ represents the tolerance for similar traits and preferences.
By averaging this degree for families in the same society, we calculated the density of cooperative families \emph{friend}$_i$ for each family $i$.}
A smaller \emph{friend} value implies that the family gained less cooperation, resulting in a decline in the growth rate, where $d_c$ represents the death rate increment due to non-cooperation.

\begin{table}[tb]
\caption{Parameters used in the model. In the results described below, the values of $b, \mu, \tau, N_f,$ and $N_s$ are fixed to those shown in the table, unless the value is described explicitly.}
  \label{table:param}
 \centering
    \begin{tabular}{l|l|c} 
Sign & Explanation & Value \\ \hline
$b$ & Intrinsic growth rate & 5.0 \\
$\mu$ & Mutation rate for $\bm{t}, \bm{p}$ & 0.1\\
$\tau$ & Tolerance for similarity & 1.0\\
$N_f$ & Initial number of families in society& 50\\
$N_s$& Number of societies in a system& 50\\
$d_c$ & Decline in mortality with cooperation& Variable\\
$d_m$ & Increase in mortality with competition& Variable\\
$\bm{t}$&Cultural traits of family& Evolve\\
$\bm{p}$&Preferences for groom traits& Evolve
    \end{tabular}
\end{table}

Next, we introduced competitive relationships with mating rivals (red dashed line in Fig. \ref{fig:scheme}(B)). Families compete with those who have similar preferences. 
\diff{The degree between family $i$ and $j$ is given by $\exp(-|\bm{p}^i-\bm{p}^j|^2/\tau^2)$.
We calculated the density of competitive families \emph{rival}$_i$ for each family $i$.} A larger \emph{rival} value implies that the family has many rivals, resulting in a decline in the growth rate, where $d_m$ represents an increase in the death rate owing to competition. Here, the strength of competition depends only on the number of families with close preferences. It is independent of the number of preferred families, because competition occurs even when there are sufficient grooms and brides \cite{Chagnon1988}.

\begin{figure*}[tb]
 \centering
  \includegraphics[width= 0.9\linewidth]{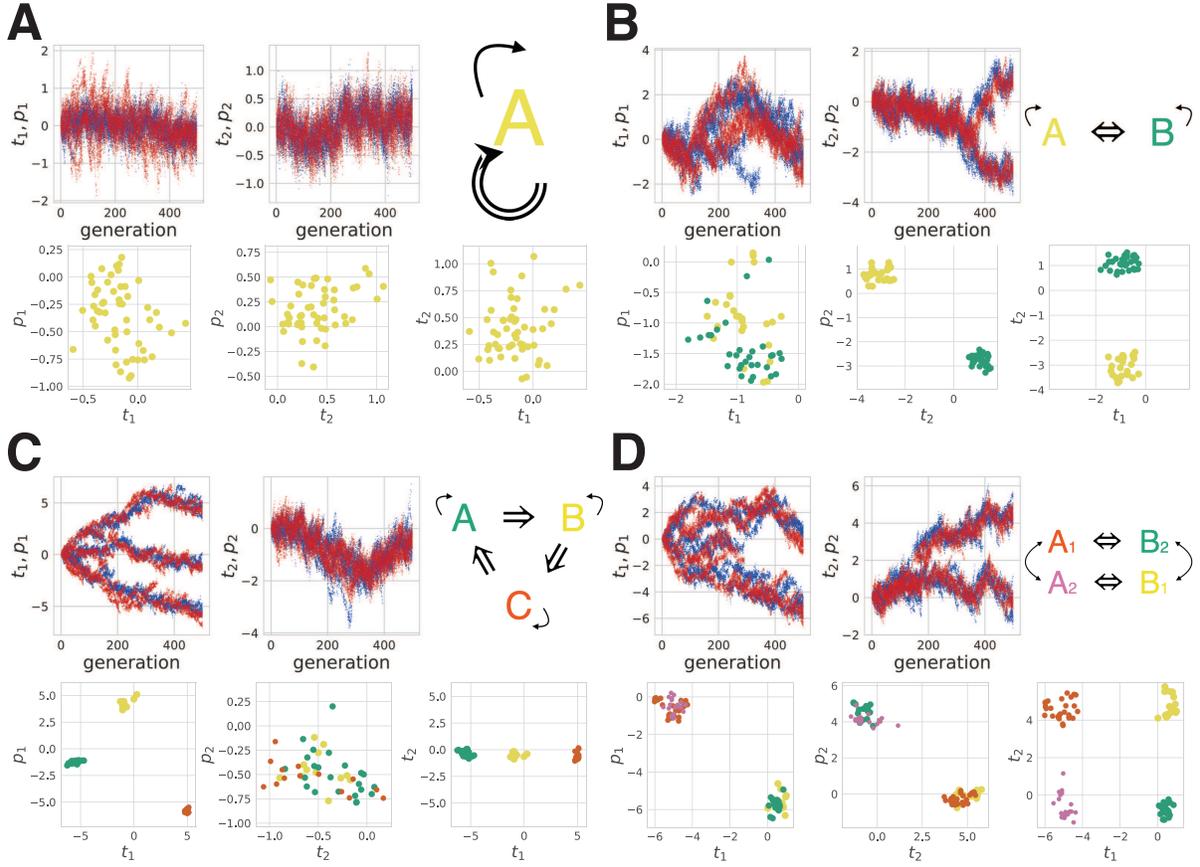}
\caption{Examples of the evolution of kinship structures. ($\bm{t}, \bm{p}$) values of families in society after 500 simulation steps. The figures show the temporal evolution of ($\bm{t}, \bm{p}$) values (upper-left), a schematic representation of the emergent structure (upper-right), and the final state (bottom). The temporal evolution of the trait and preference values of families in a society are represented in blue and red, respectively. The final states are shown as a $t_1-p_1$ map, a $t_2-p_2$ map, and a $t_1-t_2$ map, from left to right (The scales of axes differ, depending on the variance of values.)
The structures are categorised by calculating the marriage($C_m$) and descent cycles ($C_d$) as the lengths of the cycles of the flow of women and children, respectively.
(A) Incest structure without the division of clans. Marriage occurs within clan A (yellow). $d_c = 5.0, d_m = 0.1.$
(B) Dual organisation with a matrilineal descent system. Clans A (yellow) and B (green) diverge concerning the maternally inherited trait $t_2$ and prefer each other. $d_c = 0.3, d_m = 0.2.$
(C) Generalised exchange with a patrilineal descent system. Clans A (green), B (yellow), and C (orange) diverge concerning the paternally inherited trait $t_1$ and prefer others cyclically. $d_c = 0.5, d_m =  1.0.$
(D) Restricted exchange with a double descent system. Clans A$_1$ (orange), A$_2$ (pink), B$_1$ (green), and B$_2$ (yellow) exhibit pairwise marriage and descent relationships. Here, clans diverge regarding both maternally and paternally inherited traits. $d_c = 0.2, d_m =  1.0.$
}
\label{fig:emergence}
\end{figure*}

Then, we calculated the population growth as determined by the interactions of families. 
The numbers of men and women in family $i$ who survive till marriageable age are given by Poisson distribution with mean $b\exp(-d_c(1 - \emph{friend}_i) - d_m \emph{rival}_i)$, where $b$ determines the intrinsic growth rate. We adopted this form, as it is more suitable for analytical calculations.
The presented results are qualitatively independent of these specific forms. For example, $b - d_c(1 - \emph{friend}) - d_m \emph{rival}$ or $b/((1 + d_c(1-\emph{friend}))(1 + d_m\emph{rival}))$ (the latter was adopted in the previous model \cite{itao2020evolution}) essentially produces identical results if cooperation enhances, and conflict suppresses, the population. 

\diff{Finally, people get married according to their traits and preferences. The probability of marriage of men in family $i$ and women in family $j$ is proportional to $\exp(-|\bm{t}^i-\bm{p}^j|^2/\tau^2)$. After marriage, couples create their own families, bear children with inheriting traits and preferences, and then die.}

The initial values of $\bm{t}, \bm{p}$ are $(0, 0)$ in this model. Thus, at first, no rules concerning marriage or descent exist. Initially, any couple can marry, even within a nuclear family.
\diff{This assumption is set to demonstrate society-level structures that determine that the marriage rules of families can evolve, even without introducing any rules initially.
However, the results after sufficient generations are independent of the initial conditions.}
The notations and parameter values adopted in the simulations are summarised in Table \ref{table:param}.

\section*{Evolution of Kinship Structures}
The model was simulated iteratively for various parameter values listed in Table \ref{table:param}. In a simulation of 500 steps, the $(\bm{t}, \bm{p})$ values of families within a society diverged, and finally, formed some clusters in $(\bm{t}, \bm{p})$ space, as shown in Fig. \ref{fig:emergence}. 
\diff{With the pressure to increase cooperators by increasing kin and mates, isolated families without sufficient friend values are removed, and families are clustered. With the pressure to decrease mating rivals, families' preferences diverge. Accordingly, under sufficient strengths of both pressures, that is, sufficient $d_c$ and $d_m$ values, families form several discrete clusters united by marital relationships in $(\bm{t}, \bm{p})$ space.}
Siblings belonged to the same cluster. Families within the same cluster, including those who were genetically unrelated, had similar traits and recognised each other as cultural kin. They avoided marriage within their cluster and preferred mates from other clusters, that is, $\bm{t}^i\not\simeq \bm{p}^i$ to increase cooperators by acquiring mates other than their cultural kin.
Consequently, the emergent clusters were culturally united groups with the symbolic incest taboo, preferring exogamy. They can, therefore, be interpreted as clans. 
Clans were attributed based on parental traits. Here, the different clans are characterised by discretised trait values. Discretisation for $t_1$, $t_2$, or both values leads to the evolution of various descent systems.
In this model, clans’ descent relationships, as well as their marriage relationships, emerged. Here, we used the $X$-means method for clustering to optimise the number of clusters by adopting the Bayesian information criterion \cite{pelleg2000x}. The relationships between clans were determined by tracing the marriage and descent relationships of the cluster centres. The emergent structures were classified according to the cycles of marriage and descent relationships, that is, $C_m$ and $C_d$, respectively.

Various kinship structures and descent systems have evolved, as shown in Fig. \ref{fig:emergence}.
In Fig. \ref{fig:emergence}(A), only one clan, namely, A (yellow) exists and marriage occurs within it, representing an incest structure. Here, traits and preferences do not diverge.
In Fig. \ref{fig:emergence}(B), two clans, namely, A (yellow) and B (green) prefer each other (A $\Leftrightarrow$ B), representing dual organisation. In this case, traits and preferences diverge in $(t_2, p_2)$ space only. One can interpret this as a system in which maternally inherited traits $t_2$ are solely referred to for marriage and descent. Hence, a matrilineal descent system evolves. 
In Fig. \ref{fig:emergence}(C), three clans, namely, A (green), B (yellow), and C (orange) prefer other clans cyclically (A $\Rightarrow$ B $\Rightarrow$ C $\Rightarrow$ A), representing generalised exchange. 
Here, traits and preferences diverge only in $(t_1, p_1)$ space. One can interpret this as a system in which paternally inherited traits $t_1$ are solely referred to for marriage and descent. Hence, a patrilineal descent system evolves. 
\diff{Notably, in this paper, we term the system in which families choose a mate from a specific clan, as generalised exchange that is observed in some regions \cite{leach1954political, levi1969elementary}. However, the system that prohibits within-clan marriage only is not included in our model.}
In Fig. \ref{fig:emergence}(D), four clans, namely, A$_1$ (orange), A$_2$ (pink), B$_1$ (yellow), and B$_2$ (green) exhibit pairwise mating preferences (A$_1$ $\Leftrightarrow$ B$_2$ and A$_2$ $\Leftrightarrow$ B$_1$) and descent relationships (A$_1$ $\leftrightarrow$ A$_2$ and B$_1$ $\leftrightarrow$ B$_2$). Specifically, restricted exchange has evolved. Here, both maternally and paternally inherited traits significantly diverge. Hence, a double descent system evolves.

\begin{figure}[tb]
 \centering
    \includegraphics[width= 1.0\linewidth]{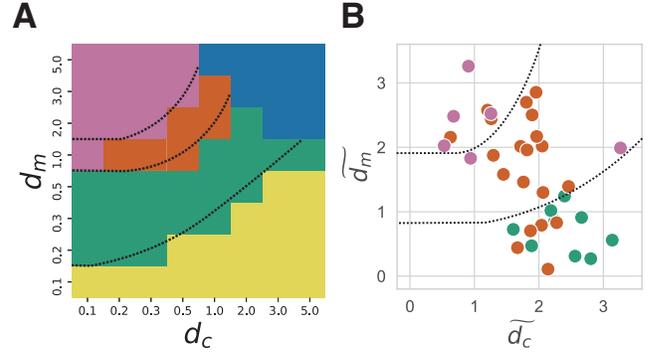}
  \caption{Phase diagrams on kinship structures. The figures show the classes of kinship structures that evolve for each environmental parameter $d_c$ and $d_m$, both theoretically and empirically.
  The incest structure is plotted in yellow, dual organisation in green, generalised exchange in orange, and restricted exchange in pink. Conditions leading to the extinction of all societies are plotted in blue.
  The dashed lines represent the rough phase boundaries of the structures. Boundaries approach $d_m / d_c = C$ asymptotically when $d_c$ is large and $d_m = C'$ when $d_c$ is small, according to the analytical calculations in the supplementary text.
    (A) Theoretical phase diagram of kinship structures. \diff{We calculated the frequencies of each kinship structure using 100 trials for each $d_c$ and $d_m$ values.} The figure illustrates the structure that evolved most frequently under each condition. Here, $N_s = N_f = 50$, and $\mu = 0.1$.
    (B) Empirical phase diagram of kinship structures, except for the incest structure. By analysing the Standard Cross-Cultural Sample (SCCS), we estimated the parameters for each society and plotted the dependencies of kinship structures on them. The estimated $\widetilde{d_c}$ and $\widetilde{d_m}$ are relative values, compared to $d_c$ and $d_m$.
    (See Fig. S5 for the empirical phase diagram of kinship structures, including the incest structure.)
    }
    \label{fig:structure_phase}
\end{figure}

\diff{Evolved kinship structures and descent systems depend on environmental parameter values $d_c$ and $d_m$. We conducted an evolutionary simulation 100 times for each condition and counted the frequencies with which each structure evolved.}
Fig. \ref{fig:structure_phase}(A) shows the dependence of kinship structures that evolved most frequently in each condition as the phase diagram. When $d_c$ far exceeded $d_m$, the incest structure (yellow) evolved most frequently. As $d_m$ increased relative to $d_c$, the emergent structure changed to dual organisation (green), generalised exchange (orange), and then to restricted exchange (pink). 
\diff{When $d_c$ is small and $d_m$ is large, societies can be composed of several endogamous clans, that is, incest structures, as shown in Fig. S2. However, it rarely occurs within the current parameter regions.}
Note that the diagram is qualitatively robust to the choice of initial conditions.

\diff{These successive transitions were accompanied by an increased number of clans within societies and a decreased probability of sustaining structures against population fluctuations. 
To estimate the phase boundary, we analytically calculated conditions for each structure to evolve. We explain it below briefly (see the supplementary text for further details).
We assume that the centres of the groom and bride clans deviate with the order of the mutation rate $\mu$ due to the fluctuations.  Because of this deviation, the degree of cooperation of the mate is reduced by the factor $\exp(-\alpha\mu^2)$ from that of the kin (where $\alpha \sim \mathcal{O}(1)$). 
Hence, for example, every family in the incest structure is kin and rival simultaneously, whereas a half is kin and rival, and the other half is mate in dual organisation. Then, recalling the above reduction, the conditions in which dual organisation is more adaptive than the incest structure are given by}
\diff{
\begin{align}
   & p_I \exp(-d_c\cdot 0- d_m \cdot 1) < \\
   &\ \ \ p_D \exp\left(-d_c\left(\frac{1}{2} - \frac{1}{2}\exp(-\alpha\mu^2)\right) - d_m \cdot \frac{1}{2}\right), \label{eq:full} \\
  \Leftrightarrow &   d_m / d_c > 1 - \exp(-\alpha\mu^2) + \frac{2}{d_c}\log p_I / p_D \label{eq:incest_dual},
\end{align}
where $p_I$ and $p_D$ denote the sustenance probability for incest structure and dual organisation, respectively (see supplementary text for their estimation). The transition to generalised or restricted exchange is estimated similarly.
In short, the transitions occur if the pressure for segmentation caused by large $d_m$ values exceeds that for clustering by $d_c$ and the relative probability for sustaining structures. Then, we derived the phase boundaries of $d_m / d_c = C$ asymptotically when $d_c$ was large and $d_m = C'$ when $d_c$ was small, as shown in Fig. \ref{fig:structure_phase}.} 

\begin{figure}[tb]
 \centering
\includegraphics[width= 1.0\linewidth]{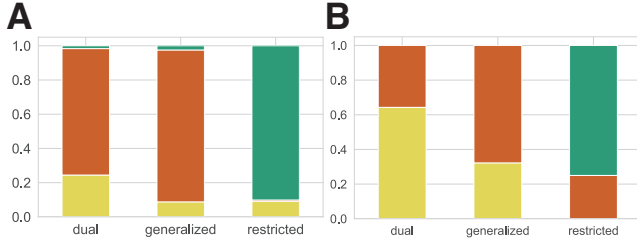}
    \caption{Frequency of descent systems for each kinship structure. Figures show the frequency of each descent system, both theoretically and empirically. The frequencies of matrilineal (yellow), patrilineal (orange), and double descent (green) systems are shown.
    (A) Theoretical phase diagram. The model was simulated by changing the $d_c$ and $d_m$ values. We calculated the frequencies of each descent system for each kinship structure. Here, $N_s = 50$, $N_f = 30$, and $\mu = 0.1$.
    (B) Empirical phase diagram. By analysing the SCCS, we identified the descent systems and kinship structures of each society. We counted the frequencies of each descent system for each kinship structure.}
    \label{fig:descent_phase}
\end{figure}

Fig. \ref{fig:descent_phase}(A) shows the dependency of descent systems on kinship structures. Double descent is dominant in restricted exchange. 
Patrilineal descent is dominant over matrilineal descent in dual organisation and generalised exchange, whereas matrilineal descent is more frequent in dual organisation.

\begin{figure}[tb]
\centering
\includegraphics[width=1.0\linewidth]{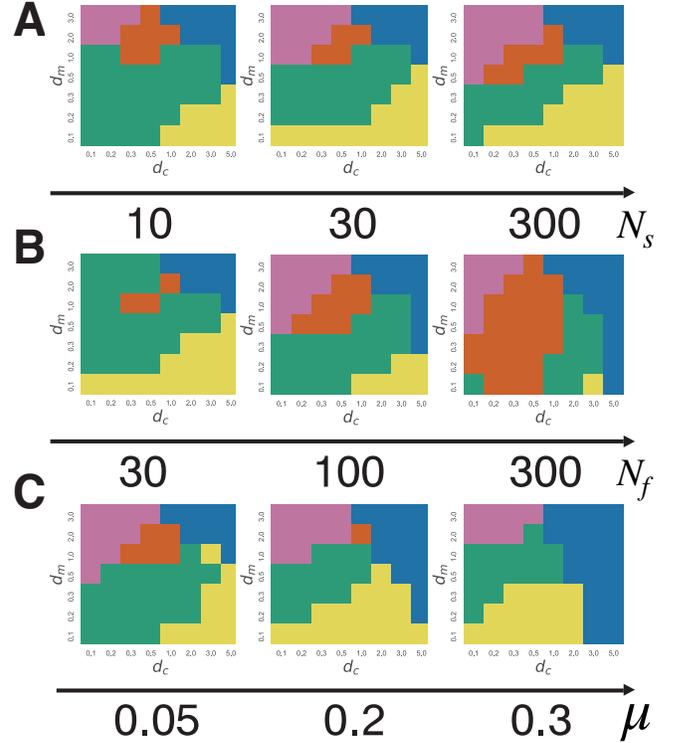}
\caption{
Dependence of phase diagrams of kinship structures on other parameters (A) the number of societies $N_s$, (B) the number of families within a society $N_f$, and (C) the mutation rate $\mu$. The incest structure is plotted in yellow, dual organisation in green, generalised exchange in orange, and restricted exchange in pink. Conditions leading to the extinction are plotted in blue. Each diagram is obtained in the same way as Fig. \ref{fig:structure_phase} (A). Unless shown on the axis, the parameter values are fixed to those in Table \ref{table:param}.}
\label{fig:kinship_phase_phase}
\end{figure}

Fig. \ref{fig:kinship_phase_phase} shows the dependence of phase diagrams of kinship structures on the number of societies in the system $N_s$, the number of families within a society $N_f$, and the mutation rate $\mu$. 
Fig. \ref{fig:kinship_phase_phase} (A) suggests that, as $N_s$ increases, restricted exchange evolves across broader parameter regions, whereas the region of dual organisation narrows. Generally, as the number of groups increases, group-level selection is strengthened in multi-level evolution \cite{itao2020evolution, takeuchi2017origin, traulsen2006evolution}. As restricted exchange requires divergence in both traits, its formation is more difficult, even for large $d_m / d_c$. Hence, group-level pressure is necessary for the evolution of such sophisticated structures. 
Next, Fig. \ref{fig:kinship_phase_phase} (B) suggests that, as $N_f$ increases, generalised and restricted exchanges evolve across broader parameter regions. Meanwhile, the incest structure and dual organisation evolve in narrower regions. 
As $N_f$ increases, fluctuations in the population of clans decrease, and thus, sophisticated structures can be easily sustained.
Realistically, however, as the population increases, interactions among families will diversify, other social organisations will evolve, and kinship structures may be destroyed. This exceeds the scope of our model.
Finally, Fig. \ref{fig:kinship_phase_phase} (C) suggests that sophisticated structures, such as restricted or generalised exchanges, disappear as $\mu$ increases. A larger $\mu$ causes larger fluctuations in traits and preferences, which can destroy more sophisticated structures.

\section*{Empirical Data Analyses}
We then verified our results on the phase diagrams of kinship structures and descent systems using the SCCS database \cite{murdock1969standard, kirby2016d}.
\diff{We classified the kinship structures of each society by identifying the composition of clans and marriage and descent rules between them. See Table S2 for further details. 
Of the 186 societies in SCCS, we identified 87 as incest structures, 14 as dual organisation, 33 as generalised exchange, and 12 as restricted exchange.
Forty societies were excluded from the analyses, as their marriage rules prohibit within clan (or family) marriage only. Fig. S3 and S4 show the geographic distributions of kinship structures and descent systems, respectively. Each structure is distributed globally, without a clear spatial pattern, suggesting that kinship structures in each region were achieved by cultural adaptation, rather than cultural transmission; this must be further investigated by phylogenetic comparative analysis.}

\begin{table}[tb]
\caption{Correlations between SCCS variables and kinship structures (excerpt).
For pairs of kinship structures, the Spearman's rank correlation between the SCCS variables and the structures was calculated. Then, the absolute values of the correlation were averaged for each pair.
We list the variables that exhibited high correlations and were relevant to $d_c$ and $d_m$, along with the average value of the correlation, and the corresponding parameters in the model.
See Table S3 for further information.
} 
  \label{table:corr}
 \centering
    \begin{tabular}{lll} 
Variable	&	Corr.	& Model	\\\hline
Tributary Payments or Taxation	&	0.58	&	$d_c$	\\
Violence against Other Ethnic Groups	&	0.57    &  $d_c$	\\
External Warfare  &	0.54		&  $d_c$	\\
Hostility towards Other Ethnic Groups	&	0.48	&  $d_c$	\\
Cross-cutting Ties	&	0.45	&  $d_c$	\\\hline
Conflict within the Society	&	0.59	&  $d_m$	\\
Violence within the Society	&	0.55	&  $d_m$	\\
Disapproval of Rape	&	0.53	&   $d_m$	\\
Disapproval of Premarital Sex	&	0.51	&  $d_m$ \\
Disapproval of Incest	&	0.41	&  $d_m$	
    \end{tabular}
\end{table}

\diff{Next, we conducted Spearman’s rank correlation analyses and calculated the correlation between SCCS variables and kinship structures. 
The database contains various variables of socio-ecological factors.
Whereas there are no variables in SCCS that exactly correspond to $d_c$ and $d_m$, $d_c$ can be related to the extent of social unity and external warfare, and $d_m$ to the attitude towards adultery and the extent of internal warfare. Notably, marriage conflict over mates arises at the family or kin group level, whereas inter-society conflict requires cooperation across different kin groups. Thus, violence within a society is related to $d_m$, and that involving other societies to $d_c$.
We calculated the correlation for each variable and listed the variables in descending order in the absolute value of the correlation. We then, found that the variables related to $d_c$ and $d_m$ were located at the top of the list (rather than the middle or bottom).} The variables that were highly correlated with kinship structures are listed in Table S3. Among them, we show the variables that can be related to $d_c$ and $d_m$ in Table \ref{table:corr}.

We estimated $d_c$ using the variables pertaining to social unity (\textit{tributary payments or taxation} and \textit{cross-cutting ties}) and society-level conflict that requires immense cooperation within society (\textit{violence against other ethnic groups}, \textit{external warfare} and \textit{hostility towards other ethnic groups}). We estimated $d_m$ using the variables pertaining to attitudes towards adultery (\textit{disapproval of rape}, \textit{disapproval of premarital sex} and \textit{disapproval of incest}) and intra-society conflict (\textit{conflict within the society} and \textit{violence within the society}).

Next, we normalised the values of each variable to set the mean $0$ and variance $1$. We changed the sign if necessary, so that larger values corresponded to larger $d_c$ or $d_m$. For some societies, the data for some variables were lacking; however, we averaged the available values to estimate $\widetilde{d_c}$ and $\widetilde{d_m}$ (hereafter, values with tilde represent those estimated by empirical data analyses). We added a constant to set the minimum values of $\widetilde{d_c}$ and $\widetilde{d_m}$ to $0$, because $d_c$ and $d_m$ were positive values in our model. 
Although the absolute magnitudes were not comparable, $\widetilde{d_c}$ and $\widetilde{d_m}$ would be positively correlated with $d_c$ and $d_m$, respectively. The empirical dependence of the kinship structures on $\widetilde{d_c}$ and $\widetilde{d_m}$ is shown in Fig. \ref{fig:structure_phase}(B). The results were qualitatively consistent with the theoretical phase diagrams for $d_c$ and $d_m$. As $\widetilde{d_m} / \widetilde{d_c} $ increased, kinship structures changed from dual organisation to generalised exchange, and then to restricted exchange. The consistency between data and model results was worse for the incest structure, as observed in Fig. S5. This may be because societies with such structures can have social systems other than kinship, regulating social unity and suppressing marital competition.

The frequency of each descent system in each kinship structure was also calculated and shown in Fig. \ref{fig:descent_phase}(B). 
The dominance of patrilineal over matrilineal descent was observed in generalised exchange. The fraction of matrilineal descent was larger for dual organisation. These are comparable with the model results, although the correspondence was much weaker than that of the kinship structures.

\section*{Discussion}
By considering cooperation among kin and mates, as well as competition among rivals in our model, we demonstrated that families formed some clusters in traits and preferences. Families within a cluster are recognised as cultural kin, and marriage occurs only among families from different clusters. Hence, the clusters of families that emerged in our model can be interpreted as clans. 
Initially, uniform traits are discretised into several clusters corresponding to distinguished clans. Furthermore, by tracing marriage and descent relationships between clans, the evolution of various kinship structures were observed. The traits and preferences were differentiated involving either paternally or maternally inherited ones only, or both. This demonstrates the evolution of patrilineal, matrilineal, and double descent systems, respectively.
Additionally, we revealed that the parameters related to $d_c$ and $d_m$ in our model can be considered as significant explanatory variables for different kinship structures, by analysing the ethnographic data of 146 societies. By estimating $d_c$ and $d_m$ from the data, we demonstrated consistency between the theoretical and empirical results of the parameter dependencies of the kinship structures and descent systems.

In cultural anthropology, ``descent theory’’ and ``alliance theory’’ have been proposed to explain kinship structures. They emphasise cooperation fostered by shared descent and marriage, respectively \cite{levi1969elementary, leach1982social}.
Here, we added the effect of marital competition. 
\diff{By introducing the evolutionary pressure to increase cooperation among kin and mates, and to decrease competition among rivals, we illustrated that diverse kinship structures evolve depending on the pressures.
Generally, it is difficult to compare historical consequences of the formation of kinship structures since chronological records are rarely available.
Nevertheless, we can explain how each structure was sustained for a specific condition.
Indeed, L{\'e}vi-Strauss demonstrated several examples of the sustenance of kinship structures. Cultural groups become divergent owing to population growth and internal conflict. Simultaneously, however, they are united by marital relationships.
Even if some of the population is damaged, structures eventually recover within several generations \cite{levi1969elementary}.}
Furthermore, we can compare our theoretical results with empirical data, and their consistency supports the plausibility of our scenario.

According to the simulations, kinship structures evolve depending on the two pressures parameterised by $d_c$ and $d_m$. That is, the importance of cooperating among kin and mates and that of avoiding marital competition determined by environmental conditions. For example, $d_c$ is related to the frequency and importance of public works or massive violence in societies, whereas $d_m$ is related to the scarcity of mates.
When the pressure for cooperation dominates the avoidance of competition, societies comprise one or several clans and most families are united as kin or mates. By contrast, as the importance of avoiding competition increases, dividing societies into more clans becomes more adaptive. Hence, as $d_m / d_c$ increases, the emergent structures change from incest structures to dual organisation, generalised exchange, and finally, to restricted exchange.
In cultural anthropology, dual organisation is categorised as the simplest form of restricted exchange, by focusing on $C_m = 2$ \cite{levi1969elementary}. Our results, however, suggest that it is closer to generalised exchange concerning environmental dependencies as expected by focusing on $C_d = 1$.

Furthermore, diverse descent systems evolved in our model. 
\diff{Under the moderate values of $d_m$, either paternally or maternally inherited traits and preferences solely diverge because of the symmetry breaking. This leads to patrilineal or matrilineal descent, respectively.} In our simulation, patrilineal descent evolved more frequently than did matrilineal. As mentioned above, the sole asymmetry of sex lies in the process of choosing a mate; that is, women (or their families) prefer certain men's traits. Thus, the selection pressure to favour those men with preferable traits leads to the divergence of paternally inherited traits.
In real-world data too, patrilineal descent is more frequent than matrilineal descent.
However, in our model, the dominance of patrilineal descent was excessive.
In some societies, grooms' families choose brides.
Furthermore, other aspects cannot be neglected.
For example, with paternal uncertainty, matrilineal descent will likely evolve \cite{holden2003matriliny}. The necessity for cooperation generally differs for men and women, depending on subsistence patterns or frequency of warfare \cite{service1962primitive}. For further discussion on the evolution of descent systems, these biases should be considered.
Nonetheless, our study shows the emergence of significant traits that are frequently inherited paternally.

In the empirical data analysis, we found the correlation between kinship structures and the status of wives, as well as $d_c$ and $d_m$ (see Table S3). Specifically, gender inequality concerning the wives' status increased in the following order: restricted exchange, dual organisation, and generalised exchange (though this cannot be directly related to the gender balance in general).
\diff{Thus, in this aspect, it may be reasonable to assume that dual organisation is more similar to restricted exchange than to generalised exchange. As the empirical data and our model exhibit, the descent system is more biased towards patrilineal descent in generalised exchange and less so in dual organisation, whereas double descent is adopted in restricted exchange. In societies with patrilineal descent systems, wives join husbands' groups after marriage \cite{service1962primitive}.
If male dominance is more frequently observed therein, we can explain the above trend. The inequality is the largest in generalised exchange; that is, between restricted exchange and dual organisation regarding environmental dependence. This suggests the benefit of analysing kinship structures to elucidate other cultural aspects of society.}

Apart from the parameters $d_c$ and $d_m$, the mutation rate $\mu$, the number of competing societies $N_s$, and the initial number of families within a society $N_f$ are also relevant in determining kinship structures (see Fig. \ref{fig:kinship_phase_phase} and Table S3). Fig. \ref{fig:kinship_phase_phase} as well as our analytical calculations suggest that sophisticated structures, such as generalized and restricted exchanges, are more fragile due to the larger fluctuation under smaller $N_s$ or $N_f$, or larger $\mu$.
Table S3 suggests that such sophisticated structures are correlated with large $N_s$ and small $N_f$.
Hence, the theoretical trend was empirically verified for $N_s$, but not for $N_f$. In reality, if $N_f$ is larger, incest structures become dominant.
This may be due to the development of social organisations other than kinship, such as political organisations, which would regulate the social relationships of families in societies with larger populations \cite{service1962primitive}. Such structurisation can be interpreted as a cultural evolutionary phenomenon; however, it is beyond the scope of our model.
\diff{Regarding $\mu$, it will be determined by how traits and preferences are inherited, and  social norms regulate precise inheritance \cite{cavalli1981cultural}.}
Thus far, however, we have been unable to estimate it from the data, and this remains a task for the future.

\diff{Our model shares some similarities with the mating preference model for sympatric speciation in biology. The evolution of several endogamous groups is shown to be a result of mating competition \cite{dieckmann1999origin} or niche construction \cite{kaneko2000sympatric}. 
Conversely, humans develop the ability of kin recognition, leading to organising the affinal network of groups by exogamy \cite{chapais2009primeval, planer2020towards}. Our model includes cooperation among mates and thus, exhibits the emergence of diverse kinship structures more than the mere divergence of groups.}

The present study has some limitations. In the model, we only focused on societies in which marriage and descent rules were strictly determined by customs. Our model concerns the elementary structures of kinship, where paternal and maternal traits are referred to independently \cite{levi1969elementary}. As population size expands, the unity of kin groups weakens, and marriage rules are relaxed, such that only marriage within the clan or nuclear family is prohibited \cite{harrell1997human}. 
\diff{Such rules to exclude unpreferable mates cannot evolve in our model. To cover observed rules comprehensively, a new model needs to be developed to consider positive and negative preferences for mates.} The descent rules also change, such that they can refer either to the father or the mother in each generation by choice, or genealogical distance only. This occurs in complex structures of kinship \cite{murdock1969standard, kirby2016d}. 

Moreover, we could only analyse the correlations between ethnographic variables and kinship structures. \diff{We could not assign the variables to $d_c$ and $d_m$ a priori. Therefore, our estimation of $\widetilde{d_c}$ and $\widetilde{d_m}$ may seem arbitrary. To measure these variables directly, it is, thus, necessary to collaborate with field studies.}
It is also desirable to conduct further analyses, such as classification learning; however, it was unfeasible in the current study owing to data insufficiency. \diff{Phylogenetic comparative analysis is also necessary to control statistical non-independence owing to shared ancestry \cite{minocher2019explaining}.} Furthermore, because of the lack of chronological data, we could not analyse the causal relationships between social structures and cultural conditions related to $d_c, d_m$, and other parameters.

Social structures, such as kinship, are formed through interactions among people over many generations. \diff{In this paper, we theoretically demonstrate such formation of macroscopic social structures through microscopic family behaviours.
It is considered difficult to explain such complex systems from basic conditions solely using simple correlation analyses  \cite{racz2020social}.}
Combined with theoretical simulations of a simple constructive model and empirical data analyses, we have demonstrated that various kinship structures emerge depending on the degrees of cooperation and avoidance of competition.
Theoretical studies, as shown here, produce explanatory scenarios by referring to empirical studies and propose relevant variables to be measured in the field. Empirical studies in the field describe notable anthropological phenomena and enable the measurement of variables to test theories. Such collaboration of theoretical and empirical studies will contribute to discussing the emergence of complex social structures and unveiling universal features in anthropology.

\section*{Method}
\subsection*{Algorithm}
To simulate population growth considering social interactions of families, the degrees of cooperation and competition were calculated by comparing trait and preference values with a tolerance parameter $\tau$. Hence, families $i$ and $j$ cooperate if $|\bm{t}^i - \bm{t}^j| / \tau$,
$|\bm{t}^i - \bm{p}^j| / \tau$, or 
$|\bm{p}^i - \bm{t}^j| / \tau$ is sufficiently small.
These conditions correspond to $i$ and $j$ being cultural kin, the women in $j$ preferring men in $i$, and the women in $i$ preferring men in $j$, respectively. Families $i$ and $j$ compete if they prefer similar families, that is, if $|\bm{p}^i-\bm{p}^j| / \tau$ is sufficiently small. Then, the possibility of marriage and the degrees of cooperation and competition were measured using a Gaussian function. For example, the degree of cooperation between cultural kin is given by $\exp(-|\bm{t}^i - \bm{t}^j|^2/\tau^2)$.

We adopted the following algorithm for population changes in families: For family $i$ and time step $n$, the numbers of unmarried men and unmarried women are denoted by $M^i(n)$ and $F^i(n)$, respectively. The intrinsic growth rate is denoted by $b$. We represented the set of families in society as $\Phi$ and the families that accept men in the family $i$ as grooms as $i'$.
The population change in family $i$ is given by

{\small
\begin{align}
d_{i, j} &= \min(|\bm{t}^i(n)-\bm{t}^j(n)|,\\
&\ \ \ \ \ \ |\bm{p}^i(n)-\bm{t}^j(n)|, |\bm{t}^i(n) - \bm{p}^j(n)|), \label{clan_eq:distance}\\
\emph{friend$^{ i}(n)$}&=\sum_{j \in \Phi} \frac{\exp(-d_{i, j}^2/\tau^2)}{\#\Phi}, \label{clan_eq:friend}\\
\emph{rival$^{ i}(n)$}&= \sum_{j \in \Phi} \frac{\exp(- |\bm{p}^i(n)-\bm{p}^j(n)|^2/\tau^2)}{\#\Phi}, \label{clan_eq:rival}\\
r &= b \exp(-d_c(1 - \emph{friend$^{ i}(n)$}) - d_m\emph{rival$^{ i}(n)$}), \label{clan_eq:fitness} \\
M^i(n) &= \text{Poisson}(r),\ F^i(n) = \text{Poisson}(r), \label{clan_eq:birth}\\
\intertext{Here, the probability for family $i$ to offer family $i'$ for marriage $P(i')$ is}
 P(i') &= \frac{\exp(- |\bm{t}^i(n)-\bm{p}^{i'}(n)|^2/\tau^2)}{\sum_{j \in \Phi} \exp(- |\bm{t}^i(n)-\bm{p}^j(n)|^2/\tau^2)}, \label{clan_eq:mate_choice}\\
\bm{t}^{i^*}(n + 1) &= (t^i_1, t^{i'}_2) + \bm{\eta},\  \bm{p}^{i^*}(n + 1) = (p^i_1, p^{i'}_2) + \bm{\eta}. \label{clan_eq:marriage}
\end{align}
}

The population growth of each family depends on \emph{friend} and \emph{rival}, as given by Eqs. (\ref{clan_eq:friend}) and (\ref{clan_eq:rival}), respectively.
The number of unmarried children in each family follows a Poisson distribution, as given by Eqs. (\ref{clan_eq:fitness}) and (\ref{clan_eq:birth}), respectively.
People are married according to the traits and preferences of their families, as shown in Eq. (\ref{clan_eq:mate_choice}). After marriage, couples give birth to children and die.  At this time, children inherit the traits and preferences of parents by adding the noise component $\bm{\eta}$ to them. This comprises two independent normal variates with mean $0$ and variance $\mu^2$ as shown in Eq. (\ref{clan_eq:marriage}).
Unmarried people can join the mating in the next step. However, those who cannot find mates within two steps die without having children.
Here, we assumed monogamy; however, the result was qualitatively independent of such a marriage system.

\subsubsection*{Data accessibility.}
Source codes for the model can be found here: \url{https://github.com/KenjiItao/clan.git}

\subsubsection*{Author contributions.}
K.I. designed the model, conducted the simulations, analysed the data, and wrote the paper. K.K. designed the model, analysed the data, and wrote the paper.

\subsubsection*{Competing interests.}
The authors declare no conflict of interest.

\subsubsection*{Funding.}
This study was partially supported by a Grant-in-Aid for Scientific Research on Innovative Areas (17H06386) from the Ministry of Education, Culture, Sports, Science, and Technology (MEXT) of Japan.

\subsubsection*{Acknowledgements.}
The authors thank Tetsuhiro S. Hatakeyama, Yuma Fujimoto, and Kenji Okubo for a stimulating discussion, and Takumi Moriyama, Koji Hukushima, Yusuke Kato, and Yasuo Ihara for illuminating comments.

\end{document}